\documentclass{amsart}
\usepackage{amsfonts}
\usepackage{graphics}

\begin{document}
\title{The velocity of dynamical chaos  during propagation of the positive Lyapunov exponents region under non-local conditions}
\author{M.N. Ovchinnikov}

\date{\today}

\begin{abstract}
The dynamics of the system is investigated when one part of the system initially behaves in a regular manner and the other in a chaotic one. The propagation of the chaos is considered as the motion of a region with the maximal Lyapunov exponent greater than zero. The time dependencies of the chaos propagation parameters were calculated for the classical and non-local models of non-stationary heat transfer. The system responses were considered to disturbances in the form of the Dirac delta function and the Heaviside step function. 

keywords: dynamic chaos, propagation velocity, Lyapunov exponent, heat conduction non-local models
\end{abstract}

\maketitle

\section{Introduction}
\label{intro}

When considering the problems of the non-equilibrium dynamics of chaotic systems, it is of interest to study the motion of the chaos-order frontier. Many articles are devoted to this problem, in which, as a rule, various aspects of the kinetic theory are analyzed \cite{Pul2016}  in the context of the stochasticity of the Boltzmann and Vlasov  equations  \cite{Fra1996}, \cite {Jab2011}, features of the disturbances and fronts propagation in non-equilibrium systems \cite{Weiss2002}, \cite{Mer2014}, \cite{Wim2003}.
Note that these problems are related to the problem of ergodicity \cite{Lee2007}, irreversibility \cite{Gas2004}, fractal of dynamic systems \cite{Gasp2004} and fluctuations of thermodynamic parameters in complex systems \cite{Jour1998}, \cite{KIN2018}, \cite{Fal1990}. 

Usually, in such systems, the maximum value of the Lyapunov exponent increases with increasing energy and becomes greater than zero when the energy reaches a certain threshold value.
In this paper did not use the method of molecular dynamics simulation, and the equations of non-stationary conductive heat transfer are analyzed. The role of the threshold value of energy, starting from which one can speak of chaotic dynamics, is played by a certain finite value of temperature.

\section{Theoretical basis}
\label{S:2}

As stated in the introduction, in this paper the motion of a region with an energy corresponding to the maximal Lyapunov exponent greater than zero into a region with an exponent value less or equal to zero. Qualitatively, the dependence of the maximum Lyapunov exponent on the energy level in the system can be represented as shown in Fig.1. Here the same figures are shown in linear and double logarithmic (internal figure) coordinates and
$E_{1}/\epsilon$ is the value of the energy (or kinetic temperature) per one degree of freedom related to the characteristic parameter of the interaction potential of particles, for example, the depth of the potential well $\epsilon$ in the Lennard-Jones potential. Usually, the threshold values of the energy $E_{ch}$, starting from which the Lyapunov exponent becomes greater than zero are small and $E_{1}/\epsilon$ is of the order of $10^{-2}$.

\begin{figure*}
\vspace*{1cm}       % Give the correct figure height in cm
\includegraphics{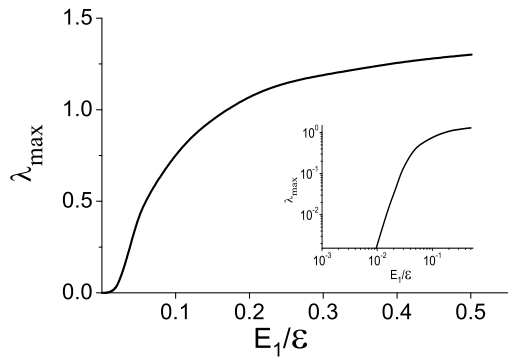}
\caption{The dependence of the Lyapunov exponent on the energy per degree of freedom.}
\label{fig:1}       % Give a unique label
\end{figure*}

Unlike \cite{MS1992}, we will study the heat transfer using the continuum mechanics equations in the one-dimensional system. In this approach, the role of the threshold energy  is played by temperature and the motion of a region with an energy value equal to $E_{ch}$ is considered as the motion of the corresponding isotherms $T=C(Constant)$.

Note that the classical heat equation
\begin{equation}
\frac{{\partial T}}{{\partial t}} = \alpha \frac{{\partial ^2 T}}{{\partial x^2 }},
\end{equation}
where $x$- is the coordinate, $T$ - the temperature, $t$ - the time, $\alpha$ - the thermal diffusivity, based on the principles of locality and local thermodynamic equilibrium and does not always adequately describe the process of unsteady heat transfer, especially at short observation times.
Also, this model has a paradox of infinite velocity of the propagation of the  disturbances.  Therefore, we will also consider the solutions of the non-local heat equations.

One of the answer for resolving the paradox of infinite perturbation velocity and the development of methods for describing non-stationary heat transfer consists in introducing non-local terms into heat equation, for example, using the concept of extended irreversible thermodynamics and the non-local Fourier law \cite{Sob1997167}, \cite{Men20101}, \cite{Cri2006}, \cite{McG2004}, \cite{Tre2004}, \cite{Cas2003}, \cite{Don2013}, \cite{Jac2016} with a transition, for example, to the telegraph type equation

\begin{equation}
\frac{{\partial T(x,t)}}{{\partial t}} + \tau \frac{{{\partial ^2}C(x,t)}}{{\partial t{}^2}} = \alpha \frac{{\partial ^2 T}}{{\partial x^2 }}.
\end{equation}
Here $\tau$ is relaxation time.

It should be said that these models are approximate as well. Following \cite{Ovc2017}, it can be assumed that the closest to reality is a model that uses “absolute randomness”  a random walk mode with discrete random jumps at a distance $\Delta x$  over time interval $\Delta t$. 
The probability density of finding the system in coordinates. For this model, the probability $P_{kn}$ of finding the system in the coordinate $x_k$ at the time moment $t_n$ can be written as \cite{Her197869}

\begin{equation}
P({x_k},{t_n}) = \frac{{({t_n}/\Delta t)!}}{{2^{\left( {{t_n}/\Delta t} \right)}}} \nonumber\\ 
\frac{{1}}{{\left( {({t_n}/\Delta t + {x_k}/\Delta x)/2} \right){\kern 1pt} {\kern 1pt} {\kern 1pt} !\left( {({t_n}/\Delta t - {x_k}/\Delta x)/2} \right)\,!\;}}  
\end{equation}
This model can lead to the telegraph equation \cite{Kac1974497}.

In this paper, it is assumed that the truth is «in the middle» between models (1-3). 
We also note that only non-local effects are considered, excluding non-linearity. In other words, thermal diffusivity, thermal conductivity, heat capacity and density are constants.

\section{The model}
\label{S:3}

Let's choose a system of units wherein  $\tau=1/2$, $\alpha = 1/2$ and continuous variables $x$ and $t$ for (3) with limits  $\Delta x\to 0, \Delta t\to 0$,  $\Delta x/\Delta t\to 1$.
Then the fundamental solutions as the response to the Dirac delta-function for (1) will be

\begin{equation}
{T^{\delta}}(G) = \frac{1}{{\sqrt {2\pi t} }}\exp ( - \frac{{{x^2}}}{{2t}}),
\end{equation}

for (2) 

\begin{equation}
{T^{\delta}}(TE) = \exp ( - t)\,{I_0}\left( {\sqrt {{t^2} - {x^2}} } \right)\,\Theta {\kern 1pt} {\kern 1pt} (t - \left| x \right|)
\end{equation}
and (3) is written as
\begin{equation}
{T^{\delta}}(RW) = \frac{{\Gamma (t + 1)}}{{\Gamma \;\left( {(t - x)/2 + 1} \right)\,\Gamma \,\left( {(t + x)/2 + 1} \right)\;{2^{t + 1}}}}.
\end{equation}
Here $\Gamma$ is Euler gamma function and (6) is the probability density distribution and can be considered as the fundamental solution for an unknown evolution equation corresponding to random walk model.

\begin{figure*}
\vspace*{1cm}       % Give the correct figure height in cm
\includegraphics{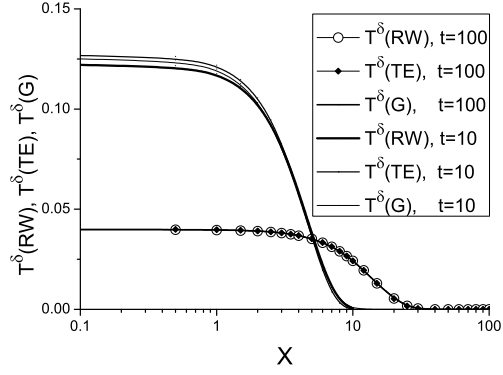}
\caption{The solutions (4-6) in time moments $t=10, 10^2$.}
\label{fig:2}       % Give a unique label
\end{figure*}

An example of fundamental solutions in the form (4-6) is shown in Fig.2  for $x>0$, $t>0$  in time moments $t = 10, 10^2$. We see that the relative differences between these solutions decrease with increasing time.

Here and below the corresponding solutions for (1) will have the additional symbol $G$ (Gauss), for (2)  $TE$ (telegraph), for (3) $RW$ (Random Walks), the $\delta$ and $\Theta$ indices the boundary conditions in the form of the Dirac delta function and Heaviside step function.

If we take the boundary conditions of the Heaviside step function $\Theta$, then the solution for (1) will be

\begin{equation}
T^{\Theta}(G) = \frac{1}{{\sqrt {2 \pi}  }}\int\limits_0^t {\frac{x}{{(  {t - t_1}  )^{3/2} }}} exp{( - \frac{{x^2 }}{{2 (t - t_{1} )}})} {\Theta(t_1)} dt_1
\end{equation}

and, using \cite {Pol2002}, for (2)

\begin{equation}
T^{\Theta}(TE) = \frac{2}{\pi }\sum\limits_{n = 1}^\infty  {\frac{1}{n}} \sin \left( {\frac{{\pi nx}}{L}} \right)\left[ 1 - \exp ( - t){\kern 1pt} \cos (Kt) -K^{-1/2} \exp ( - t){\kern 1pt} \sin (Kt) \right].
\end{equation}
Here $K=\sqrt{ (\pi ^2 n^2 /L^2 - 1) }$.
Solutions $T^{\Theta}(G)$, $T^{\Theta}(TE)$ for $x>0$, $t>0$ with the initial condition $T(x,0)=0$ and the boundary conditions $T(0,t) = \Theta(t)$  and $T(L,t) = 0$ are shown in Figure 3 in time moments $t=10, 10^2, 10^3, 10^4$. Here $x <L$, $L$ is the final length of the system for calculations. We see that solutions become very close to each other with increasing time (respectively,  $t/ \tau >> 1$ and $t/\Delta{t} >> 1$).

\begin{figure*}
\vspace*{1cm}       % Give the correct figure height in cm
\includegraphics{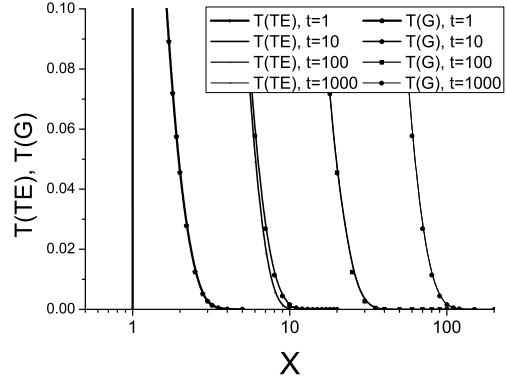}
\caption{The solutions (7,9-10) in time moments $t=10, 10^2, 10^3,10^4$.}
\label{fig:3}       % Give a unique label
\end{figure*}

\section{ Results }
\label{S:4}

Let us consider solutions (4-6) and calculate the isotherm coordinates $X_C^{\delta ,G}$ at arbitrary time moments for the corresponding constant temperatures $T = C$. One of the finite value of $T$ will be a threshold borderline of  chaos.

For the solution (4), such a coordinate $X_C^{\delta ,G}$ will be
\begin{equation}
X_C^{\delta ,G}  = \sqrt {2t\ln (1 / {\sqrt {2\pi t C^2 }}}  
\end{equation}
under condition

\begin{equation}
t < {1 / (2\pi C^2 )},
\end{equation}
for the solution (5)

\begin{equation}
X_C^{\delta ,TE} { \approx \sqrt {2t\ln \frac{1}{{\sqrt {2\pi tC^2 } }}} \left( {1 - \frac{1}{t}(\beta_1 \ln t + \beta_2 )} \right)},
\end{equation}
and for the solution (6)

\begin{equation}
X_C^{\delta ,RW}  \approx \sqrt {2t\ln \frac{1}{{\sqrt {2\pi tC^2 } }}} \left( {1 - \frac{1}{t}(\beta _3 \ln t + \beta_4 )} \right).
\end{equation}

Accordingly, the velocities $V_C^{\delta ,G}$, $V_C^{\delta ,TE}$, $V_C^{\delta ,RW}$ of these isotherms movement  will be for (4)
\begin{equation}
V_C^{\delta ,G}  = {{\left( {\ln \frac{1}{{\sqrt {2\pi tC^2 } }} - \frac{1}{2}} \right)} \mathord{\left/
 {\vphantom {{\left( {\ln \frac{1}{{\sqrt {2\pi tC^2 } }} - \frac{1}{2}} \right)} {\sqrt {2t\ln \frac{1}{{\sqrt {2\pi tC^2 } }}} }}} \right.
 \kern-\nulldelimiterspace} {\sqrt {2t\ln \frac{1}{{\sqrt {2\pi tC^2 } }}} }}
\end{equation}
for (5)
(\begin{equation}
V_C^{\delta ,TE}  \approx \frac{{\left( {\ln \frac{1}{{\sqrt {2\pi tC^2 } }} - \frac{1}{2}} \right)}}{{\sqrt {2t\ln \frac{1}{{\sqrt {2\pi tC^2 } }}} }}\left( {1 - \frac{{\beta _1 \ln t + \beta _2 }}{t}} \right) + \sqrt {2t\ln \frac{1}{{\sqrt {2\pi tC^2 } }}} \left( {\frac{{\beta _1 \ln t + \beta _2 }}{{t^2 }} - \frac{{\beta _1 }}{{t^2 }}} \right)
\end{equation}13),
and
\begin{equation}
V_C^{\delta ,RW}  \approx \frac{{\left( {\ln \frac{1}{{\sqrt {2\pi tC^2 } }} - \frac{1}{2}} \right)}}{{\sqrt {2t\ln \frac{1}{{\sqrt {2\pi tC^2 } }}} }}\left( {1 - \frac{{\beta _3 \ln t + \beta _4 }}{t}} \right) + \sqrt {2t\ln \frac{1}{{\sqrt {2\pi tC^2 } }}} \left( {\frac{{\beta _3 \ln t + \beta _4 }}{{t^2 }} - \frac{{\beta _3 }}{{t^2 }}} \right)
\end{equation}
for (6). Here $t>>\beta_1 ln(t)$, $\beta_1 =-0.143$, $\beta_2 =-0.286($lnC$)-0.64$, $\beta_3 =0.0656+0.724($lnC$)$, $\beta_4 =-0.1917(lnC)-0.8087$ for $10^{-5}<C<0.4$.

Similarly, for (7) with the boundary condition in the form of $\theta$-function
the $X_C^{G,\Theta }$, $X_C^{TE,\Theta }$  coordinates for (7) will be defined as
\begin{equation}
X_C^{G,\Theta }  = \gamma_1 \sqrt t 
\end{equation}
and for (8) as

\begin{equation}
X_C^{TE,\Theta }  \approx \gamma_1 \sqrt t \left( {1 - \frac{\gamma_2}{t} + 0(t^{ - 2} )} \right),
\end{equation}
were $\gamma_1 =-0.0146($lnC$)^2-0.4906 ($lnC$)+0.6014$, $\gamma_2=-0.0201($lnC$)^2-0.3843 ($lnC$)-0.4848$, and velocities, respectively, as

\begin{equation}
V_C^{G,\Theta }  = \frac{\gamma_1 }{{2\sqrt t }}
\end{equation} 
and
\begin{equation}
V_C^{TE,\Theta }  \approx \frac{\gamma_1 }{{2\sqrt t }}\left( {1 + \frac{\gamma_2}{t}} \right).
\end{equation}

Taking into account the existing of fluctuations in real systems and the infinite time of chaos formation, we can introduce the concept of the width of the zone of chaos formation in time moment $t$, for example, in the form
\begin{equation}
\Delta X_c^{G,\Theta }  = \frac{\gamma_1 }{{2\sqrt t }} {\Delta T_{ch}}
\end{equation}
for (18) and

\begin{equation}
\Delta X_C^{TE,\Theta } \approx \frac{\gamma_1 }{{2\sqrt t }}\left( {1 + \frac{\gamma_2}{t}} \right)\Delta T_{ch}
\end{equation}
for (19). Here $\Delta{T}_{ch}$ is the characteristic time of chaos formation.

\section{ Basic conclusions  }
\label{S:5}

The propagation of the region with chaotic behavior is considered above as the motion in the dynamic system of a region with a maximal Lyapunov exponent greater than zero. This process is associated with the energy redistribution using continuum mechanics model. Under this assumption, it is actual to know the temperature level that exceeds a certain finite threshold value, starting from which chaotic dynamics arise in the system.
Since the temperature redistribution can be described by various models, this study considers both the classical and non-local conductive heat equation and two types of boundary conditions were chosen: in the form of the Dirac delta function and the Heaviside step function.
	
For the $\delta$-function type initial perturbation, the velocity of chaos propagation can be estimated by the formula (13) for equation (1). For models (2-3), deviations from this solution decrease with time approximately as $ln{t}/ t$ (14-15) and faster.

For a $\theta$-type Heaviside boundary condition the velocity of chaos propagation can be estimated by the formula (18) proportional $1/t$ for model (1). For model (2), the deviations from this solution decrease with time are approximately (19) as $1/t$.
Thus, significant differences in the behavior of the systems under consideration (1-3) are observed at times $t <10^2$ in the selected system of units, which corresponds, for example, to Lennard-Jones systems $t <10^{-9}$s.

In this study, the goal was to consider the differences between models (1-3) in terms of the non-locality influence. Note that in real conditions it is also necessary to take into account the nonlinear dependencies of the thermodynamic   parameters. We also note that in chaos propagation, obviously, there are zones of chaos formation, so, the chaos-order borderline will be blurred.

An example of an experimental observation of the motion of a chaos-order borderline is the propagation of a paramagnetic-ferromagnetic boundary starting to ferromagnetic heating.

\address{
Kazan Federal University, Kazan, Russia\\
\email{marov514@gmail.com}\\
}

\end{document}